\documentclass[NJP]{iopart}
\usepackage{graphicx}
\usepackage{amssymb}
\usepackage{breqn}
\usepackage{cases}
\usepackage{iopams}
\usepackage{hyperref}
\hypersetup{
  colorlinks=true,
  allcolors=blue
}

\begin{document}

\title{Quasiclassical approach to synergic synchrotron-Cherenkov radiation in polarized vacuum}

\date{\today}
\author{I.~I.~Artemenko}
\author{E.~N.~Nerush}
\ead{nerush@appl.sci-nnov.ru}
\author{I.~Yu.~Kostyukov}
\address{Institute of Applied Physics of the Russian Academy of
Sciences, 46 Ulyanov St., Nizhny Novgorod 603950, Russia}

\begin{abstract}
    The photon emission by an ultrarelativistic charged particle in extremely strong magnetic field
    is analyzed, with vacuum polarization and photon recoil taken into account. The vacuum
    polarization is treated phenomenologically via refractive index. The photon emission occurs in
    the synergic (cooperative) synchrotron-Cherenkov process [J.~Schwinger, W.~Tsai and T.~Erber, Annals
    of Physics, 96 303 (1976)] which is similar to the synchrotron emission rather than to the Cherenkov
    one. For electrons, the effect of the vacuum polarization on the emission spectrum is not
    evident even beyond the probable onset of non-perturbative quantum electrodynamics (QED).
    However, the effect of the vacuum polarization on the emission spectrum can be observable for
    muons already at $\gamma B / B_S \approx 30$, with $\gamma$ the muon Lorentz factor, $B$ the
    magnetic field strength and $B_S$ the critical QED field. Nevertheless, vacuum
    polarization leads to only $10\%$ enhancement of the maximum of the radiation spectrum.
\end{abstract}

\maketitle

\section{Introduction}
\label{introduction}

Quantum electrodynamics (QED) predicts nonlinear dielectric properties of the vacuum in strong
magnetic field caused by virtual electron-positron pairs. The Kramers--Kronig relations connect
vacuum refractive index with pair photoproduction probability, and the latter have been studied in
strong crystalline fields~\cite{Uggerhoj05} and in laser field~\cite{Burke97, Bamber99}. Still
direct experimental evidence of vacuum refractive index is absent, and many set-ups have been
proposed to detect and measure it, e.g. x-ray diffraction on a double-slit formed by two
counterpropagating intense laser pulses~\cite{King10, King10a, King14}, or changes in polarization
of x or gamma photons due to vacuum birefrigence in strong laser field~\cite{Heinzl06, Ilderton16,
Nakamiya17, Bragin17}. The idea behind these proposals is not only to measure vacuum refractive
index but to test QED in a not-yet-investigated region of extreme laser fields. Moreover,
investigation of vacuum polarization becomes important in the light of Ritus--Narozhny conjecture
of perturbative QED breakdown at certain conditions~\cite{Fedotov17, Yakimenko19, Baumann19,
Blackburn19b, Piazza19a}.

The fields of intensity $10^{23} \mathrm{-} 10^{24}~\mathrm{ W} \, \mathrm{cm}^{-2}$ is expected
in near future thanks to facilities such as ELI-NP~\cite{ELINP},
ELI-beamlines~\cite{Garrec14}, Apollon~\cite{Apollon}, Vulcan 2020~\cite{Vulcan2020},
XCELS~\cite{Bashinov14} and others. Therefore, the field of the order of $10^{-3} \times B_S$ will be
available which results the vacuum refractive index $n$ such that $\delta n = n - 1 \sim 10^{-10}$
for photons with energy $\lesssim 1~\mathrm{GeV}$~\cite{McDonald86}, with $B_S = m^2 c^3 / e
\hbar$ the Sauter--Schwinger critical QED field~\cite{Fedotov10}, $m$ and $e > 0$ the electron mass
and charge magnitude, respectively, $c$ the speed of light and $\hbar$ the reduced Plank constant.
Despite such small value of $\delta n$, the Lorentz factor $\gamma \sim 10^5$, available for
electrons nowadays, is enough to reach the speed of a charged particle greater than the phase speed of
the photons, hence the Cherenkov emission may occur. Such estimates drives the recent interest to
Cherenkov emission in the polarized vacuum~\cite{Dremin02, Macleod19, Bulanov19b}. However, the
results of these papers should be reconsidered because of simplified approach used there. A charged
particle in the strong field inevitably moves along a curved trajectory that prevents plain
Cherenkov radiation. The trajectory
curvature determines the
radiation formation length and is crucial for the emission process. Furthermore, there are unified
emission process~\cite{Schwinger76}, and it is not possible to distinguish "Cherenkov" and "Compton" emission
mechanisms in the considered situation, as Refs.~\cite{Dremin02, Macleod19} does.
Ref.~\cite{Bulanov19b}, although considers Cherenkov emission and nonlinear Compton scattering as a
single process, uses expression for the emitted energy and for the formation length [equations~(16)
and (17) wherein] as if the particle moves along the straight line and emits photons in a plain
Cherenkov process.  At the same time, earlier works on the considered topic contain not only
qualitative estimates, but expressions for the spectrum and for the photon emission probability.

In 1966 Erber was the first who pointed out the possibility of Cherenkov radiation in polarized
vacuum~\cite{Erber66}. He used expression for pair photoproduction in a constant magnetic field and
dispersion relation to compute the real part of the refractive index, following work of Toll in
1952, see reference~\cite{Tsai75} and references therein. In 1969 Ritus considered possibility of
Cherenkov radiation in a constant crossed electromagnetic field~\cite{Ritus70}, using photon
Green's function obtained year before by Narozhny. Thus, there is no need in two laser pulses
which create regions with pure magnetic field, and it is enough to use single laser pulse to
induce vacuum polarization. Then, in 1976 Schwinger, Tsai and Erber with QED
mass operator method obtained the general expression for the spectrum of the photon emission by
a charged particle which moves both in a constant magnetic field and in a medium with $n \neq
1$~\cite{Schwinger76}. They pointed out that "there is actually only a single emission act,
synergic synchrotron-Čerenkov radiation, for which a correspondence with either Čerenkov emission
or synchrotron radiation can be established only in the respective limits of vanishing field or
matter density", and that "the practical import of this synergism is that the radiation depends
sensitively on both positive and negative values of $n - 1$". They demonstrated~\cite{Schwinger76,
Erber76} that depending on parameters, both amplification and suppression (quenching) of the photon
emission may occur. Finally, synergic
synchrotron-Cherenkov radiation in gases was observed in the experiment~\cite{Bonin86}, which
results agree well with the analysis which treats Cherenkov and synchrotron radiation as limiting
manifestations of a unified process.

Another interesting result of Erber {\it et al.} is that the Cherenkov condition for electrons $v
> c / n$ (with $v$ the electron velocity) is not enough for spectrum of the synchrotron-Cherenkov
radiation in polarized vacuum to be different from purely synchrotron spectrum. The sufficient condition
for this occurs extremely strict \{see equations~(8.8e) and~(8.11) in Ref.~\cite{Erber76}\}:
\begin{equation}
    \label{chi_Erber76}
    \chi = \frac{B}{B_S} \gamma \gtrsim 10^5,
\end{equation}
which for $B / B_S \sim 10^{-3}$ yields enormous energy $mc^2 \gamma \sim 100~\mathrm{TeV}$. Here
$B$ is the magnetic field strength, $\gamma$ the electron Lorentz factor, and the electron velocity
$\mathbf{ v }$ is assumed to be perpendicular to the magnetic field. Erber then suppose that the
condition~\eref{chi_Erber76} indicates that higher order QED corrections besides vacuum polarization
should be also taken into account, i.e. QED is no longer a perturbative theory. Indeed the
threshold $\chi$ value~(\ref{chi_Erber76}) is even far beyond the conjectured value of the
perturbative QED breakdown~\cite{Fedotov17, Ritus70} $\chi \sim 1 / \alpha^{3/2} \approx 1.6 \times
10^3$.

The aim of the current paper is manifold. First, the physical picture of photon emission by
ultrarelativistic particles is recalled and applied to the synchrotron emission in a medium with
$\delta n \ll 1$, within the classical theory (section~\ref{classical_theory}). A special attention
is paid to the synchronism between the emitting particle and the emitted wave. Second, the general
quantum formulas for spectral and angular distribution of the emitted photons in synergic
synchrotron-Cherenkov process are obtained (section~\ref{quantum_theory}), for that  quasiclassical
theory of Baier and Katkov~\cite{Baier98} is used. This allows to take into account photon recoil
neglected in Refs.~\cite{Schwinger76, Erber76}. Third, in section~\ref{estimates} the onset of
Cherenkov corrections to the synchrotron spectrum is found. Following the proposal of
Erber~\cite{Erber76}, synchrotron-Cherenkov emission by particles heavier than electrons is
considered in details in section~\ref{estimates_muons}. It is shown that the onset of Cherenkov
corrections to the synchrotron spectrum for muons occurs at much lower value of $\chi$ than that
for the electrons,
due to enlarged formation length and weakened photon recoil.
Section~\ref{conclusion} is the conclusion.

\section{Photon emission by ultrarelativistic particle in classical theory}
\label{classical_theory}

Calculation of the radiation of ultrarelativistic charged particle for $n = 1$ can be found in many
textbooks, e.g. in~\cite{Jackson62}. However, these calculations are often difficult to tailor to
the case $n \neq 1$. Here the general formulas for angular and spectral distributions of the
emitted energy are recalled and applied to the synchrotron radiation. Despite $n = 1$ is used in
this section, the approach used here allows obvious generalization to the case $n \neq 1$, if
$\delta n \ll 1$.

\subsection{General formulas}

For the sake of simplicity one can consider emission of electromagnetic waves by a current density
$\mathbf{j}$ inside a virtual superconductive rectangular box (resonator or cavity) of size $ L_x \times L_y
\times L_z $.  The emitted field can be decomposed by complex resonator modes with well-known
sine-cosine spatial and $ \exp(-i \omega_s t) $ temporal structure:
 \begin{equation}
     \label{decomposition}
     \mathbf{E} = \sum_s C_s \mathbf{E}_s, \qquad
     \mathbf{B} = \sum_s C_s \mathbf{B}_s,
 \end{equation}
 where $s$ is the generalized mode number and $\omega_s$ is the mode cyclic frequency, $\mathbf{E}$
 and $\mathbf{B}$ are the electric and magnetic field, respectively.  The modes can be chosen
 orthogonal, with the following normalization:
 \begin{equation}
     \label{norm}
     \frac{1}{8 \pi} \int_V \left( \mathbf{E}_s \mathbf{E}_l^* + \mathbf{B}_s \mathbf{B}_l^*
     \right) \, dV = \hbar \omega_s \delta_{sl},
 \end{equation}
 where the symbol $^*$ means complex conjecture. Hence the energy of the emitted field is
 \begin{equation}
     I = \frac{1}{8 \pi} \int_V \left( \mathbf{EE}^* + \mathbf{BB}^* \right) \, dV = \sum_s \hbar
     \omega_s |C_s|^2,
 \end{equation}
 and $ |C_s|^2 $ can be interpreted as the emission probability of the photon of mode $ s $.

 To find $C_s$, one can start from Maxwell's equations:
\begin{eqnarray}
    \label{rotE}
    \nabla \times \mathbf E = -\frac{1}{c}\frac{\partial \mathbf B}{\partial t}, \\
    \label{rotB}
    \nabla \times \mathbf B =  \frac{1}{c}\frac{\partial \mathbf E}{\partial t} + \frac{4 \pi}{c}
    \mathbf j, \\
    \nabla \mathbf E = 4 \pi \rho, \qquad
    \nabla \mathbf B = 0,
\end{eqnarray}
with $\rho$ and $\mathbf j$ are the charge and the current density, respectively.  Let the current
$\mathbf j$ emits during $t \in (t_1, t_2)$, and $\mathbf j = 0$, $\rho = 0$ for $t < t_1$ and $t >
t_2$. Thus, the decomposition~\eref{decomposition} is valid for $t > t_2$.  One can multiply
Eq.~\eref{rotE} on $\mathbf B_s^*$, and subtract it from Eq.~\eref{rotB} multiplied on $\mathbf
E_s^*$. Then the result can be integrated over the space and time that yields
\begin{equation}
\eqalign{
    \label{jE}
    \left.
        \int_V \left( \mathbf{E_s^* E + B_s^* B} \right) \, dV \;
    \right\vert_{t_1}^{t_2} +
    4 \pi \int_{t_1}^{t_2} \int_V \mathbf j \mathbf E_s^* \, dV \, dt \\
    = c   \int_{t_1}^{t_2} \oint_S \left(\mathbf B \times \mathbf E_s^* -
                                 \mathbf E \times \mathbf B_s^*
                                 \right) \, dS \, dt,
}
\end{equation}
with $V$ a volume of the virtual box and $S$ its boundary.

The cavity can be chosen
big enough such that $\mathbf{E = B} = 0$ at the boundary, in this case the right-hand side of
equation~\eref{jE} is zero. Furthermore, $\mathbf E = \mathbf B = 0$ at $t = t_1$,
hence from equation~\eref{jE}, taking into account equations~\eref{decomposition} and \eref{norm},
one gets
\begin{equation}
    \label{c_s}
    C_s = -\frac{1}{2 \hbar \omega_s} \int_t \int_V \mathbf{j E}_s^* \, dV \, dt.
\end{equation}

Equation~\eref{c_s} has clear physical meaning. Being multiplied by $\hbar \omega_s C_s^*$,
it expresses the equality between the energy emitted into the mode $s$, and work of the current
$\mathbf j$ over the one-half field of the emitted mode. This work peaks if there is a
synchronism between the current and the field of the mode. Note also that formula~\eref{c_s} is
similar to one for the amplitude of an oscillator driven by an external force.

 For an ultrarelativistic electron, which emits mostly in the forward direction, the computation
 of $ C_s $ can be further simplified. First, the current of the electron is
 \begin{equation}
    \mathbf{j} = -e \mathbf{v} \delta(\mathbf{r} - \mathbf{r}(t)),
 \end{equation}
with $\mathbf{r}(t)$ the electron position.  Second, each of the complex modes is formed by eight
complex plane waves $\propto \exp(-i \omega_s t + i \mathbf{k}_s \mathbf{r})$ (except a few modes
with wave vector parallel to the box boundaries). This yields eight terms in the integral over $t$
in equation~\eref{c_s}. It can be noted that one of the terms oscillates much slower than the others
which hence can be dropped \{e.g., if $k_x \approx \omega_s / c$ and $x(t) \approx c t $, then
$\exp[ i \omega t - i k_x x(t)]$ cannot be dropped, whereas $ \exp[ i \omega t + i k_x x(t)] $ can
be\}. Let the remaining term corresponds to a wave with the polarization direction $\mathbf{e_s}$ (with $e_s^2 =
1$). The amplitude of this remaining wave, $a_s$, can be found from the normalization~\eref{norm}:
the wave energy is $ \hbar \omega_s / 8 $ hence $ a_s = (2 \pi \hbar \omega_s / V)^{1/2} $.
Therefore,
\begin{equation}
    \label{c_s_electron}
    C_s = - \frac{e}{2} \sqrt{ \frac{\pi}{\hbar \omega_s V} }
    \int_t \mathbf{v e}_s
    \exp \left[ i \omega_s t - i \mathbf{k}_s \mathbf{r}(t) \right] \, dt.
\end{equation}

An ultrarelativistic particle emits photons in a narrow cone around the direction of the particle
velocity. Thus the energy radiated in a certain direction can be readily computed from the energy
of the modes. The density of the modes which has a plane-wave component in some certain unit solid
angle and unit frequency interval can be found from the boundary conditions for the virtual
superconducting box. From this, the full emitted energy can be expressed using the energy radiated
per unit frequency interval and per unit solid angle:
\begin{equation}
    \label{I}
\eqalign{
    I = \sum_s \hbar \omega_s |C_s|^2
    = \frac{\hbar V }{\pi^3 c^3} \int \int \omega^3 \sum_{\mathbf e} |C_{\mathbf e}|^2
                                      \, d\omega \, d\Omega \\
    = \frac{e^2}{4 \pi^2 c^3} \int \int \omega^2 \sum_{\mathbf e}
   \left|
       \int \mathbf{e} \mathbf{v}(t)
       \exp \left[ i \omega t - i \mathbf{k r}(t) \right] \, dt
   \right|^2 \, d\omega \, d\Omega,
}
\end{equation}
with $ \mathbf{e}_i$ ($i = 1, 2$) the polarization directions. Equation~\eref{I} is very useful in
estimating the radiation timescales and the radiation formation length, that discussed for the
sinchrotron radiation in the next section, and for the synchrotron-Cherenkov radiation
in section~\ref{rf_length}.

\subsection{Synchrotron emission and the timescales}

\begin{figure*}
	\includegraphics{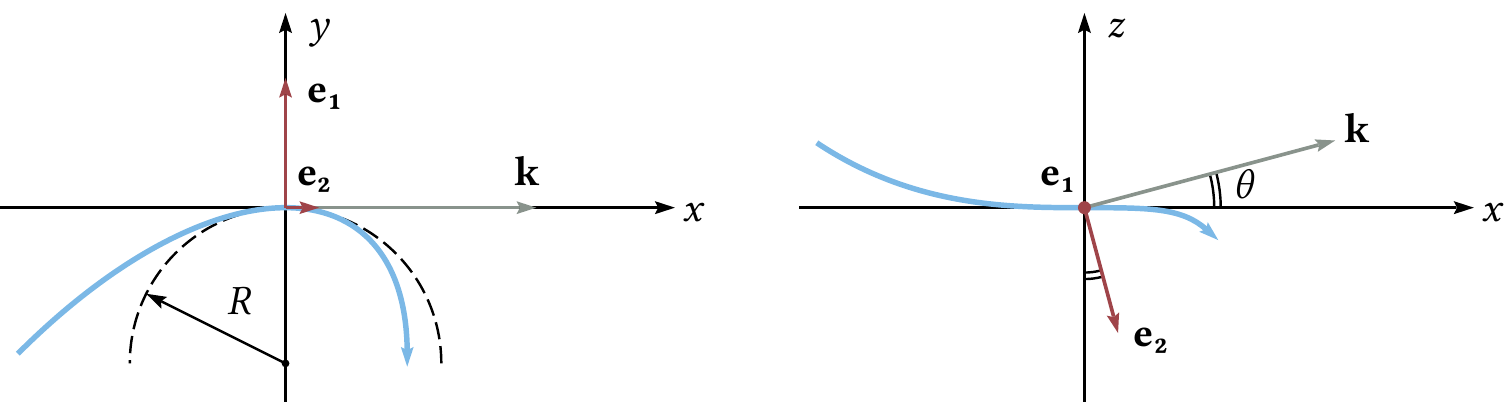}
    \caption{\label{schematic} Local coordinates used in the computations. For a given wave number
    $\mathbf k$ the origin is the point on the electron trajectory (thick blue line) where $\mathbf
    k$ is perpendicular to the normal vector of the trajectory. Thus, the $x$ axis is tangent to
    the trajectory, the $y$ axis is parallel to the normal vector (hence the $xy$ plane is the
    osculating plane), and the $z$ axis is chosen by the right-hand rule. The polarization vector
    $\mathbf e_1$ is chosen to be on the $y$ axis, and $\mathbf e_2$ to be perpendicular to
    $\mathbf e_1$ and $\mathbf k$.}
\end{figure*}

The key feature of the photon emission by an ultrarelativistic particle is the synchronism between
the particle and the emitted wave, as seen from equations~\eref{c_s}, \eref{c_s_electron} and
\eref{I}. The phase of the exponent in these equations in the case $n =
1$ (hence $k = \omega / c$), varies slowly in vicinity of the point where the angle between
$\mathbf v$ and $\mathbf k$ is minimal. For
the sake of simplicity we assume that there is only one such point, and it is in the origin of the
local coordinates~(figure~\ref{schematic}), and the particle is in the origin at $t = 0$.

In the synchrotron approximation, or local-constant-field approximation, the particle trajectory is
described locally like a circular orbit fully determined by the local curvature radius $R$ and the
Lorentz factor $\gamma$:
\begin{eqnarray}
    x \approx R \sin(vt / R) \approx vt - \frac{(vt)^3}{6 R^2}, \\
    y \approx R \left[ \cos(vt / R) - 1 \right] \approx -\frac{(v t)^2}{2 R}.
\end{eqnarray}
Then, the pre-exponential functions in the integrand of equation~\eref{I} can be written as
follows:
\begin{eqnarray}
    \label{ve1}
    \mathbf{ve}_1 \approx c^2 t / R, \\
    \mathbf{ve}_2 \approx c \sin \theta, \\
    \label{exp}
    \exp [i \omega t - i \mathbf{kr}(t)] \approx \exp[i \phi(t)],
\end{eqnarray}
where the phase $\phi$ contains only linear and cubic terms:
\begin{equation}
    \label{synchrotron_phi}
    \phi(t) = 2 \pi \left[\frac{t}{\tau_\parallel} + \left( \frac{t}{\tau_\perp}
    \right)^3 \right].
\end{equation}
Here $\tau_\parallel$ and $\tau_\perp$ are the timescales of dephasing between the electron and the
emitted wave caused by the longitudinal (along the $x$ axis) and transverse (along the $y$ axis)
electron motion, respectively:
\begin{eqnarray}
    \label{synchrotron_tau_parallel}
    \tau_\parallel = \frac{4 \pi}{\omega (\theta^2 + 1 / \gamma^2)}, \\
    \label{synchrotron_tau_perp}
    \tau_\perp = \left(\frac{12 \pi \gamma^2}{\omega \, \omega_B^2}\right)^{1/3}.
\end{eqnarray}
where the effective magnetic field strength $B$ is introduced for convenience such that $v =
\omega_B R / \gamma$, and
\begin{equation}
    \omega_B = \frac{eB}{mc}
\end{equation}
is the cyclotron frequency in this field. Note that $\tau_\parallel$ depends on $v$ (hence on
$\gamma$) and does not depend on $R$, whereas $\tau_\perp$ depends on $R \propto \gamma / \omega_B$
and does not depend separately on $\gamma$.

Well-known equations~(14.78) and (14.83) from the textbook~\cite{Jackson62} which describe angular and
spectral distribution of the synchrotron photons can be easily got from equations~\eref{I} and
\eref{ve1}-\eref{synchrotron_phi}. The key feature of the synchrotron
spectrum is the critical frequency~\cite{Jackson62}
\begin{equation}
    \omega_c = \frac{3 \gamma^3 c}{R} = 3 \omega_B \gamma^2.
\end{equation}
The energy emitted per unit frequency interval per unit solid angle, $d^2 I / d\omega d\Omega$, has
maximum at $\theta = 0$ and $\omega \approx 0.42 \times \omega_c$.

If $\omega \gg \omega_c$ or
$\theta \gg 1 / \gamma$, the emitted energy sharply tends to zero, that can be explained
with $\tau_\parallel$ and $\tau_\perp$.
The critical frequency corresponds to $\tau_\perp / \tau_\parallel = 3 / (4
\pi)^{2/3} \approx 0.56$. If $\omega$ increases beyond $\omega_c$, or if $\omega \sim \omega_c$ and
$\theta$ increases beyond $1/\gamma$, then $\tau_\parallel$ becomes smaller than $\tau_\perp$.
Hence, the exponent~\eref{exp}-\eref{synchrotron_phi} oscillates strongly hence the synchrotron
integrals
tend to zero.  As shown in the next sections, the presence of the refractive index, spin
contribution and photon recoil change the basic equation for the photon emission probability.
However, the emission probability is still governed by the synchronism between the emitted wave and
the electron, hence, by $\tau_\parallel$ and $\tau_\perp$, though equations for them should be corrected.

\section{Synchrotron-Cherenkov radiation in quantum electrodynamics}
\label{quantum_theory}

\subsection{Quasiclassical theory of the synchrotron-Cherenkov radiation}

In order to take into account the refractive index $n = 1 + \delta n$ (which is assumed
close to unity, $|\delta n| \ll 1$) in the classical formula~\eref{I}, one should not change
anything, except the relation between the photon frequency and the wave vector in the phase,
\begin{equation}
    k = n \omega / c.
\end{equation}
The mode structure, the energy of the modes and their normalization can be taken unchanged
in the case $|\delta n| \ll 1$.  This situation replicates in QED. If one follows the Baier--Katkov quasiclassical
derivation of the spectral and angular distribution of the synchrotron photons~\cite{Baier98}, he/she finds
that the presence of the refractive index changes nothing in it except the phase in the
exponential. However, to isolate the phase one should reorganize the quasiclassical
formula~\cite{Baier98} \{see also, for example, equation~(6) in the
supplementary material of Ref.~\cite{Wistisen18}\}:
\begin{equation}
    \label{from_BKS}
    \eqalign{
    \frac{d^2 I}{d\omega d\Omega} = \frac{e^2}{4 \pi^2 c} \left\{
        \frac{ \varepsilon^2 + {\varepsilon'}^2 }{ 2 \varepsilon^2 } \right. \\
        \left. \times
        \left| \int dt\, \frac{ \mathbf{n \times [(n - \bbeta) \times \dot {\bbeta}]} }
                              { (1 - \mathbf{n \bbeta})^2 }
               \exp [ i \omega' ( t - \mathbf{n \brho} ) ]
        \right|^2 \right. \\
        \left.
      + \frac{1}{2} \left(\frac{ \hbar \omega m c^2 }{ \varepsilon^2 } \right)^2
        \left| \int dt\, \frac{ \mathbf{n \dot {\bbeta}} }
                              { (1 - \mathbf{n \bbeta})^2 }
               \exp [ i \omega' ( t - \mathbf{n \brho} ) ]
        \right|^2
    \right\}
    }
\end{equation}
with $\bbeta = \mathbf{v} / c$, $\brho = \mathbf{r}/c$, $ \mathbf{n} = c \mathbf{k} / \omega $
(still $|\mathbf n| = 1$ here), and
\begin{eqnarray}
    \varepsilon' = \varepsilon - \hbar \omega, \\
    \omega' = \omega \varepsilon / \varepsilon'.
\end{eqnarray}
One can note that
\begin{eqnarray}
    \frac{d}{dt} \frac{1}{1 - \mathbf{n \bbeta}}
      = \frac{ \mathbf{n \dot {\bbeta}} }{ (1 - \mathbf{n \bbeta})^2 }, \\
    \frac{d}{dt} \frac{ \mathbf{ n \times [ n \times \bbeta ] } }{ 1 - \mathbf{n \bbeta} }
      = \frac{ \mathbf{ n \times [(n - \bbeta) \times \dot {\bbeta}] } }{ (1 - \mathbf{n \bbeta})^2 }, \\
    \frac{d}{dt} \exp [ i \omega' ( t - \mathbf{n \brho} ) ]
      = i \omega' (1 - \mathbf{n \bbeta}) \exp [ i \omega' ( t - \mathbf{n \brho} ) ].
\end{eqnarray}
Hence, integrating by parts one gets
\begin{equation}
    \eqalign{
    \label{bks_I}
    \frac{d^2 I}{d\omega d\Omega} = \frac{e^2 {\omega'}^2}{4 \pi^2 c} \left\{
        \frac{ \varepsilon^2 + {\varepsilon'}^2 }{ 2 \varepsilon^2 }
        \sum_{\mathbf e}
        \left| \int dt\, \bbeta \mathbf{e}
               \exp [ i \omega' ( t - \mathbf{n \brho} ) ]
        \right|^2 \right. \\
        \left.
      + \frac{1}{2} \left(\frac{ \hbar \omega m c^2 }{ \varepsilon^2 } \right)^2
        \left| \int dt\,
               \exp [ i \omega' ( t - \mathbf{n \brho} ) ]
        \right|^2
    \right\},
    }
\end{equation}
where the product $ \mathbf{n \times [n \times \bbeta]} $ is rewritten using $ \bbeta \mathbf{e}_1
$ and $ \mathbf{\bbeta e}_2 $. Now, to take into account the refractive index $n$, one should
set $|\mathbf n| \equiv |c\mathbf k / \omega| = n$ in the exponential functions in
equation~\eref{bks_I}.

Equation~\eref{bks_I} differs from the classical one~\eref{I} by two quantum features. First, an
additional "spin" term appears in~\eref{bks_I} \{the last term, which originates from the spin
flips~\cite{Kirsebom01, Berestetskii82}\}. Second, the radiation recoil arises, which is reflected,
first of all, in the fact that $\omega$ is substituted with a higher frequency $\omega'$ in the
exponential.
Hence, if the photon energy is about the electron energy, the synchronism is strongly affected by
the recoil. Particularly, the recoil effect squeezes the photon spectrum such that it is limited by
the energy of the electron.

All the terms in equation~\eref{bks_I}, including the spin term, contain the same phase in the
exponent, $\exp [ i \omega' ( t - \mathbf{n \brho} ) ] \approx \exp [i\phi(t)]$ with
\begin{eqnarray}
    \label{bks_phi}
    \phi(t) = 2 \pi \left[\frac{\varsigma t}{\tau_\parallel} + \left( \frac{t}{\tau_\perp}
    \right)^3 \right], \\
    \varsigma = \mathrm{sgn}( \theta^2 + 1/\gamma^2 - 2 \delta n).
\end{eqnarray}
This means that, as for the classical synchrotron emission, the spectrum of the synchrotron-Cherenkov
emission at the given frequency $\omega$ is governed by the only two timescales 
\begin{eqnarray}
    \label{bks_tau_parallel}
    \tau_\parallel = \frac{4 \pi}{\omega' |\theta^2 + 1 / \gamma^2 - 2 \delta n|}, \\
    \label{bks_tau_perp}
    \tau_\perp = \left(\frac{12 \pi \gamma^2}{\omega' \, \omega_B^2}\right)^{1/3}.
\end{eqnarray}
which though differs from the timescales of the classical synchrotron spectrum given by
equations~\eref{synchrotron_tau_parallel} and \eref{synchrotron_tau_perp}.

The refractive index brings a novel effect: the sign of the linear term can be changed, i.e. if
Cherenkov condition is fulfilled, $\beta n > 1$, then at least for $\theta \approx 0$ one has
$\varsigma = -1$. In the subsequent sections~\ref{rf_length} and \ref{estimates} it is demonstrated
that in the case of $\varsigma < 0$ the emission spectrum can differ dramatically from the synchrotron
one. In the remaining part of this section the effect of the radiation recoil is
considered in detail, because of its importance both for the case $\varsigma = +1$ and $\varsigma =
-1$.

Similarly to the classical synchrotron emission, the critical frequency $\omega_c$ can be
introduced
\begin{eqnarray}
    \label{bks_omega_c}
    \omega_c = \frac{\varepsilon \omega'_c}{\varepsilon + \hbar \omega'_c}, \\
    \label{bks_omega'_c}
    \omega'_c = \frac{3 \omega_B \gamma^2}{|1 - 2 \delta n \gamma^2|^{3/2}}.
\end{eqnarray}
such that for it $\tau_\perp$ and $\tau_\parallel$ are of the same order for $\theta = 0$, namely
for $\omega = \omega_c$ one has $\tau_\perp / \tau_\parallel = 3 / (4 \pi)^{2/3} \approx 0.56$. 

If the quantum parameter
is small, $\chi = \gamma B / B_S \ll 1$, quantum formulas tend to classical ones, i.e. radiation recoil is
negligible, $\omega_c \approx \omega_c' \approx 3 \omega_B \gamma^2$, and the spin term is
negligible. In this case (if additionally $\delta n = 0$), the maximum of $d^2 I /d\omega d\Omega$
is in the point $\theta = 0$ and $\omega \approx 0.42 \omega_c$ that reveals the physical meaning
of $\omega_c$ in this case.

In the quantum limit, $\chi \gg 1$ (and for $\delta n = 0$), equation~\eref{bks_omega'_c} yields
$\hbar \omega'_c \gg \varepsilon$, that looks non-physical if one neglects the effect of radiation
recoil and sets $\omega_c = \omega'_c$.  Actually, the radiation recoil changes significantly the
critical frequency, and $\omega_c$ differs significantly from $\omega'_c$: $\hbar \omega_c \approx
\varepsilon [1 - 1 / (3 \chi)]$. Thus $\omega_c$ is very close to the upper spectrum bound $\varepsilon
/ \hbar$. Therefore, in the quantum limit almost for all frequencies one has $\tau_\perp
\ll \tau_\parallel$, and the term $t / \tau_\parallel$ can be neglected in the phase of the
exponential.

For the refractive index of the polarized vacuum the Cherenkov condition is
fulfilled only in the quantum case, i.e. $n \beta > 1$ can be reached only if $\chi
\gg 1$ (this is discussed in section~\ref{introduction} and especially in section \ref{estimates}). As seen from
equations~\eref{bks_tau_parallel} and \eref{bks_tau_perp}, the refractive index affects $\tau_\parallel$ only, hence only the
linear term in the phase. However,  in the
quantum case this term is negligible for almost all frequencies in the synchrotron spectrum.
Therefore, in order to modify the synchrotron spectrum, the refractive index should be large enough
not only to change $\tau_\parallel$, but make it much lower than in the case of $\delta n = 0$.  That
is why the Cherenkov condition is far from being enough to change the synchrotron
spectrum.

\subsection{Radiation formation length}
\label{rf_length}

The radiation formation length is the length of the electron path which contributes most to
the integrals in equations~\eref{I} and \eref{bks_I}. Obviously, the radiation formation
length depends on the frequency of the emitted wave, although for synchrotron emission often $\omega
\sim \omega_c$ is assumed.

The radiation formation time (i.e. the radiation formation length divided by $c$) can be estimated
by consideration of the following integral:
\begin{equation}
    \mathcal{I}(t_a, t_b) = \int_{t_a}^{t_b} f(t) \sin [\phi(t)] \, dt,
\end{equation}
with $f(t)$ and $\phi(t)$ slowly varying functions and $\sin [\phi(t)]$ contains many oscillation periods
on the interval $[t_a, t_b]$. The contribution of a single oscillation period can be estimated as
follows:
\begin{equation}
    \label{I_t0_t2}
\eqalign{
    \mathcal{I}(t_0, t_2) = \int_{t_0}^{t_2} f(t) \sin [\phi(t)] \, dt \\
    \approx 2 f T |_{(t_0 + t_1) /2} - 2 f T |_{(t_1 + t_2) /2}
    \approx -\int_{t_0}^{t_2} \left( \frac{d}{dt} fT \right) \,dt.
}
\end{equation}
Here $t_0$, $t_1$, $t_2$ are time instants which correspond to $\phi = 0, \; \pi, \; 2 \pi$,
respectively. Continuous function $T(t)$ is approximately equal to the local period of function $\sin
[\phi(t)]$.  Obviously, the estimate~\eref{I_t0_t2} for $I(t_a, t_b)$ can be extended to an
arbitrary
integer number of periods between $t_a$ and $t_b$. In this case the integral can be estimated
as the difference between the integral of the first "bump" (the first half-period) and the last one,
whereas the intermediate "bumps" do not contribute. Finally, the integral which determines the convergence
speed of $\mathcal{I}(-\infty, \infty)$ can be estimated as follows:
\begin{equation}
    \label{convergence_speed}
    \mathcal{I}(t, \infty) \sim f(t) T(t),
\end{equation}
where we assume that $\lim_{t \to \infty} fT = 0$.

The local oscillation period for the phase~\eref{bks_phi}  far from the saddle points is
\begin{equation}
    \label{T_estimate}
    T(t) \approx 2 \pi \left( \frac{d \phi}{dt} \right)^{-1}
    = \left(\frac{\varsigma}{\tau_\parallel} + \frac{3 t^2}{\tau_\perp^3} \right)^{-1}.
\end{equation}
The integrals in equation~\eref{bks_I} contain $f(t) = 1$ and $f(t) = t$, both lead to the same
radiation formation time, thus for simplicity we set $f(t) = t$ from here on.

\begin{figure*}
	\includegraphics{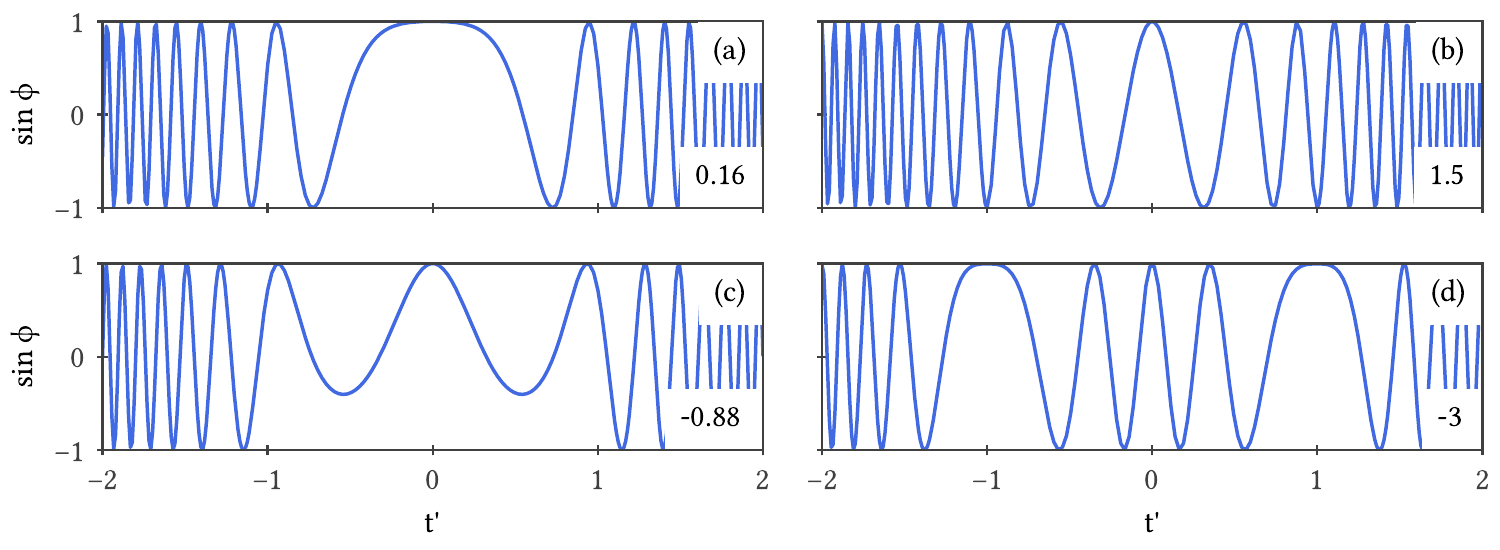}
    \caption{\label{sinus} Function $\cos [\phi(t')]$ for the phase given by
    equation~\eref{synchrotron_phi}, namely $\phi(t') = 2 \pi (a t' + {t'}^3)$ with $t' = t /
    \tau_\perp$ and $a = \varsigma \tau_\perp / \tau_\parallel$. Proper values of $a$ are given in
    the lower-right corners.}
\end{figure*}

For synchrotron emission of low frequencies ($\tau_\perp \lesssim \tau_\parallel$, $\varsigma = +1$)
the leading "bump" of the integrand has width about $\tau_\perp$ [see figure~\ref{sinus}(a)],
and for $t \gg \tau_\perp$ one gets $\mathcal{I}(t, \infty) \propto 1 / t$. Hence the radiation
formation time is $t_{rf} \sim \tau_\perp$ in this case. For high frequencies [$\tau_\perp \gtrsim
\tau_\parallel$, $\varsigma = +1$, see figure~\ref{sinus}(b)] the integrand contribution decays,
$\mathcal{I}(t, \infty)
\propto 1 / t$, only if $t \gg t_s$ with
\begin{equation}
    t_s = \sqrt{ \frac{\tau_\perp^3}{3 \tau_\parallel} }
\end{equation}
a point where the linear and the cubic terms in the phase yield the same oscillation periods. Thus, here
the radiation formation time is $t_{rf} = t_s \gtrsim
\tau_\perp \gtrsim \tau_\parallel$.

A special consideration needed for the Cherenkov branch of the synchrotron-Cherenkov
emission ($\varsigma = -1$). If $\tau_\perp \lesssim \tau_\parallel$, the sign of the linear term
is unimportant and $t_{rf} \sim \tau_\perp$. However, in the case $\tau_\perp \gtrsim
\tau_\parallel$ the integrand changes significantly [see figures~\ref{sinus}(c) and (d)]: most
contribution to the synchrotron-Cherenkov integrals comes from the regions around two saddle points
$t = \pm t_s$.  If $\tau_\parallel \ll \tau_\perp$, the phase near a saddle point (say, $t = +t_s$)
can be approximated with a parabolic dependency:
\begin{equation}
    \phi = 2 \pi \left( -\frac{t}{\tau_\parallel} + \frac{t^3}{\tau_\perp^3} \right)
    \approx \frac{6 \pi t_s}{\tau_\perp^3} (t - t_s)^2 + \mathrm{const},
\end{equation}
where the term $2 \pi (t - t_s)^3 / \tau_\perp^3$ is neglected at the right-hand-side. Then the
width of the bump at $t_s$ can
be found,
\begin{equation}
    T(t_s) \sim \left( \frac{\tau_\perp^3}{t_s} \right)^{1/2} \sim \tau_\perp^{3/4}
    \tau_\parallel^{1/4},
\end{equation}
Hence, the cubic term is actually small: $(t - t_s)^3 / \tau_\perp^3 \sim T^3(t_s) / \tau_\perp^3
\sim (\tau_\parallel / \tau_\perp)^{3/4} \ll 1$. The parabolic phase dependency leads to quite fast
convergence of the integrals, e.g. $fT \sim 1 / |t - t_s|$ for $T(t_s) \ll |t - t_s| \ll t_s$.

An important consecuence of the estimations above is that $\tau_\parallel \ll T(t_s) \ll
\tau_\perp \ll t_s$ in the case $\varsigma = -1$ and $\tau_\parallel \ll \tau_\perp$. This means
that the saddle points $t = \pm t_s$ are far away from each other, and there is almost random
phase shift between the integrals around these points. Thus, instead of coherent sum of the
integrals one should sum resulting probabilities which are computed separately for $t = +t_s$ and
$t = -t_s$ points. This also means that the radiation formation length in this case should be
estimated not as the distance between these points, but as the width of the leading bumps, $t_{rf}
\sim T(t_s) \sim \tau_\perp^{3/4} \tau_\parallel^{1/4} \ll \tau_\perp$.

Summing up the above, the radiation formation length for synchrotron-Cherenkov radiation is
\begin{equation}
    \label{t_rf}
    t_{rf} / \tau_\perp \sim
      \cases{
          \eqalign{
          1 \qquad
              & \mathrm{for} \quad \tau_\perp \lesssim \tau_\parallel, \\
          (\tau_\perp / \tau_\parallel)^{1/2} \qquad
              & \mathrm{for} \quad \tau_\perp \gtrsim  \tau_\parallel, \; \varsigma = 1, \\
          (\tau_\parallel / \tau_\perp)^{1/4} \qquad
              & \mathrm{for} \quad \tau_\perp \gtrsim  \tau_\parallel, \; \varsigma = -1.
          }
      }
\end{equation}
Besides the radiation formation length, the synchrotron-Cherenkov integrals depend on the ratio $\tau_\perp
/ \tau_\parallel$ itself. For instance, one can note that the integrals decay exponentially with
the increase of $\tau_\perp / \tau_\parallel$, if $\tau_\perp \gg  \tau_\parallel$ and $\varsigma =
+1$ [see figure~\ref{sinus}(b)]. The emission probability becomes
negligible in this case. Taking this into account, one notes from equation~\eref{t_rf} that in all
the noticeable cases the radiation formation length is less
or about $\tau_\perp$, which do not depend on the refractive index. Note also that in the regime of
seemingly dominance of Cherenkov radiation, $\delta n \gg 1 / \gamma^2$, one has $\varsigma = -1$
and $\tau_\parallel \ll \tau_\perp$ [see figure~\ref{sinus}(d)], and the radiation formation time is small, $t_{rf} \ll
\tau_\perp$. This differs significantly from the case of plain Cherenkov radiation, where the
radiation formation length can be extremely large.

\begin{figure*}
	\includegraphics{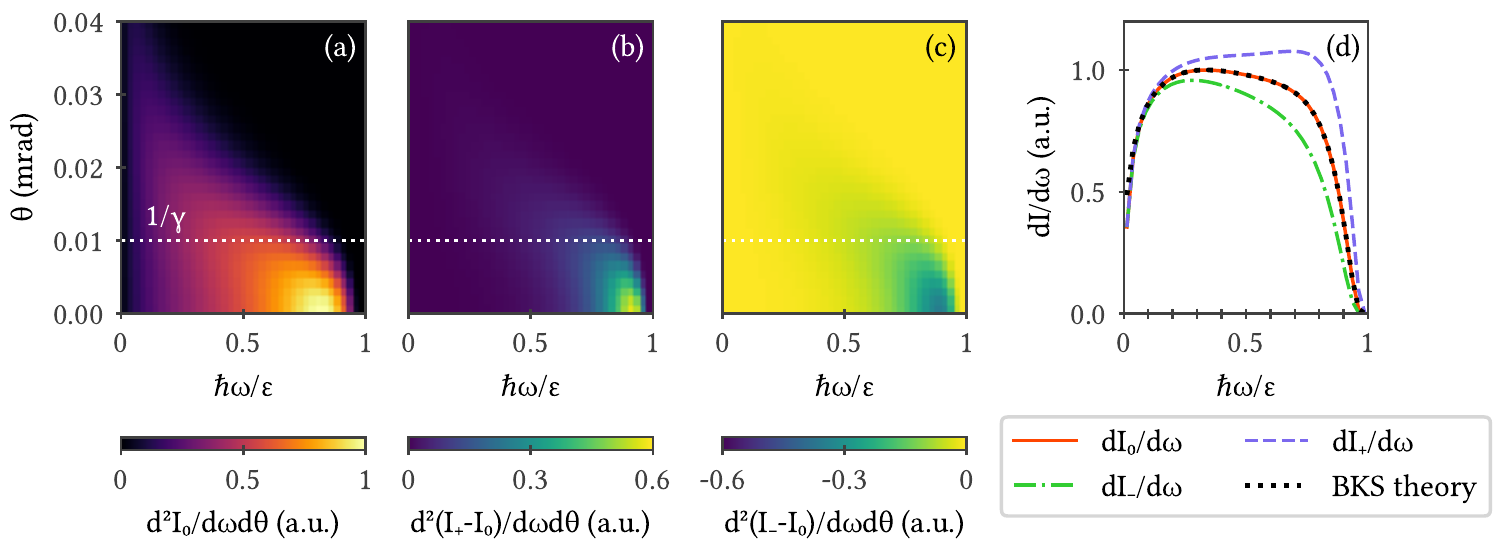}
    \caption{\label{small_n} The effect of the refractive index on the spectrum of the quantum
    synchrotron radiation in the case of relatively small $\delta n$. The energy emitted by an electron
    per unit photon frequency and per unit angle $\theta$: (a) $d^2 I_0 / d\omega d\theta$, for
    $n = 1$ ($\delta n = 0$), (b) $d^2 I_+ / d\omega d\theta$, for
    $\delta n > 0$, and (c) $d^2 I_- / d\omega d\theta$, for
    $\delta n < 0$. (d) The spectrum of the synchrotron emission
    ($\delta n = 0$) computed (dotted line) analytically and (solid line) numerically, as well as
    the spectra of the synchrotron-Cherenkov radiation for (dashed line) $\delta n > 0$ and (dash-dotted
    line) $\delta n < 0$, computed numerically. In all the cases $\varsigma = +1$ for the whole
    frequency and angular range. See text for details.}
\end{figure*}

The presence of the refractive index with $\delta n \neq 0$ can however significantly influence the emission probability.
The effect of the refractive index in the case $\varsigma = +1$ is shown in figure~\ref{small_n},
where $I_0$, $I_+$ and $I_-$ are the emitted energy for $n = 1$, $n = 1 + 0.1 / \gamma^2$ and $n =
1 - 0.1 / \gamma^2$, respectively.  Figures~\ref{small_n}(a)-(c) show frequency and angular
distribution, whereas figure~\ref{small_n}(d) depicts the energy emitted per unit frequency interval.
Distributions $d^2 I / d\omega d\theta$ for figures~\ref{small_n}(a)-(c) are computed numerically
as described in section~\ref{numerical_implementation}, for $\gamma = 1 \times 10^5$ and $b = 3
\times 10^{-5}$ (hence $\chi = 3$). Lines in figure~\ref{small_n}(d) are computed by summing up
these distributions along the $\theta$ axis, except the black dotted line which correspond to
the analytical expression for the pure synchrotron emission ($n \equiv 1$) found in the framework of Baier--Katkov--Strackovenko
theory~\cite{Baier98, Berestetskii82}. 

In the high-frequency region, $\omega \gtrsim \omega_c$, the timescales relates as $\tau_\perp \gtrsim \tau_\parallel$ in
the absence of $\delta n$. Thus, as the presence of the refractive index with $\delta n > 0$ leads
to the increase of $\tau_\parallel$, it also leads to the substantial increase of the emission
probability [compare figure~\ref{sinus}(b) and figure~\ref{sinus}(a)]. In the opposite case of
refractive index with $\delta n < 0$, the timescale $\tau_\parallel$ decreases that quenches the
emission probability. The described picture is confirmed well by figure~\ref{small_n}, where the
critical frequency is $\hbar \omega_c = 0.9 \times \varepsilon$.

In the low-frequency region ($\omega \ll \omega_c$ hence $\tau_\perp \ll \tau_\parallel$) the
sinchrotron integrals do not depend on $\tau_\parallel$ which the only depend on the refractive
index. Therefore, the presence of the refractive index with $\delta n \neq 0$ does not change the
low-frequency spectrum, as seen in figure~\ref{small_n}.
If one increases the parameter $\chi$, the critical frequency tends to
$\varepsilon / \hbar$, and the region where the effect of the refractive index is noticeable
becomes narrow and
pinned to $\varepsilon / \hbar$.

\begin{figure*}
	\includegraphics{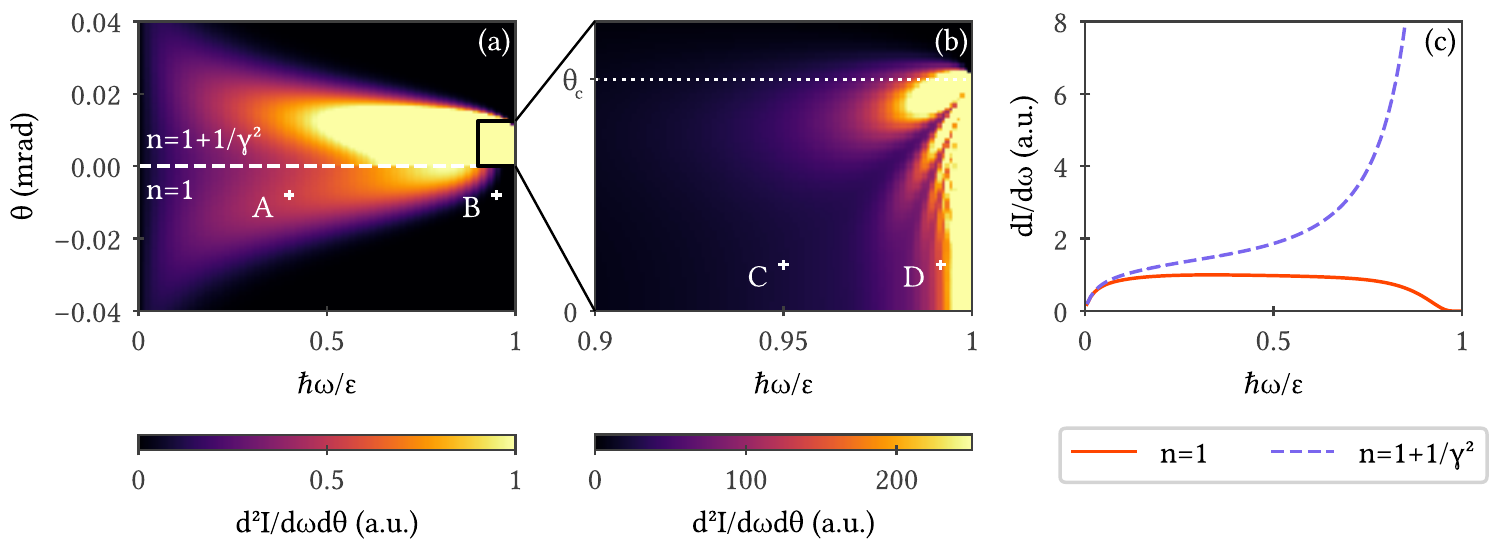}
    \caption{\label{sinus2} The effect of the refractive index on the spectrum of the synchrotron
    radiation in the case $\delta n > 1 / (2 \gamma^2)$. The energy emitted by an electron per unit
    photon frequency and per unit angle $\theta$ for (a, lower half) pure synchrotron emission with
    $\delta n = 0$, and for (a, upper half; b) synchrotron-Cherenkov emission with $\delta n =
    1/\gamma^2$. (c) Radiation spectrum for (solid line) $\delta n = 0$ and (dashed line) $\delta n
    = 1 / \gamma^2$ cases. The Cherenkov angle $\theta_C = (2 \delta n - 1/\gamma^2)^{1/2} =
    0.01~\mathrm{mrad}$ is shown by dotted white line in (b). Points A, B, C and D in (a, b)
    correspond to phase dependency shown in figures~\ref{sinus} (a), (b), (c) and (d),
    respectively. See text for further details.}
\end{figure*}

The effect of the refractive index becomes more dramatic if the Cherenkov condition is fulfilled,
$\delta n > 1 / (2 \gamma^2)$. Such case for $\delta n = 1 / \gamma^2$ is shown in
figure~\ref{sinus2} for $\gamma = 1 \times 10^5$ and $b = 3 \times 10^{-5}$ (hence $\chi = 3$ and
$\hbar \omega_c = 0.9 \times \varepsilon$). For the chosen refractive index and $\theta = 0$,
$\tau_\parallel$ remains the same as for the case $\delta n = 0$, but the sign of the linear term
in the phase~\eref{bks_phi} changes, $\varsigma = -1$. As before, the low-frequency part of the
spectrum remains the same in the both cases, $\delta n = 0$ [lower half of figure~\ref{sinus2}(a)
and solid red curve in figure~\ref{sinus2}(c)] and $\delta n > 0$ [upper half of
figure~\ref{sinus2}(a) and dashed blue curve in figure~\ref{sinus2}(c)]. However, in the case
$\delta n > 0$ the high-frequency part of the spectrum is extremely enhanced, such that the overall
emitted energy is more than an order of magnitude greater than in the case $\delta n = 0$.

Points A, B, C and D in figures~\ref{sinus2}(a, b) exactly correspond to $a \equiv \varsigma
\tau_\perp / \tau_\parallel$ used in figures~\ref{sinus} (a), (b), (c) and (d), respectively. The
fine structure in the radiation distribution at high frequencies is seen in figure~\ref{sinus2}(b),
which shows in details the rectangular region of figure~\ref{sinus2}(a) marked with black solid
line (note the different color scale in these figures). This fine structure emerges due to
the interference of the contributions yielded by the two bumps seen in figure~\ref{sinus}(d).

Figure~\ref{sinus2} looks quite encouraging, however, the refractive index of the polarized
vacuum depend on the photon frequency, and have both $\delta n > 0$ and $\delta n < 0$ parts. The
latter corresponds to high photon energies. This, together with the fact
that the high-frequency region (where refractive index influence the radiation spectrum) becomes
extremely narrow in the case $\chi \gg 1$, makes almost impossible to reveal the effect of vacuum
polarization on the synchrotron radiation, at least for electrons (or positrons). The radiation
spectrum for electrons is discussed in details in
section~\ref{estimates}, whereas the next section, section~\ref{numerical_implementation}, is
devoted to details of numerical computation of the spectrum.

\subsection{Numerical implementation}
\label{numerical_implementation}

Similarly to the transition from equation~\eref{c_s_electron} to equation~\eref{I}, one can find
from $d^2 I / d\omega d\Omega$ [equation~\eref{bks_I}] the photon emission probability summed up
for both polarizations, $W = \sum_{\mathbf{e}} |C_{\mathbf{e}}|^2$, which is more convenient for
numerical simulations:
\begin{equation}
    \eqalign{
    \label{W}
    W = \frac{e^2 c^2 \pi {\omega'}^2}{4 \hbar \omega^3 V} \left\{
        \frac{ (\varepsilon^2 + {\varepsilon'}^2) }{ 2 \varepsilon^2 }
        \sum_{\mathbf e}
        \left| \int dt\, \bbeta \mathbf{e}
               \exp [ i \omega' ( t - \mathbf{n} \brho ) ]
        \right|^2 \right. \\
        \left.
      + \frac{1}{2} \left( \frac{ \hbar \omega m c^2 }{ \varepsilon^2 } \right)^2
        \left| \int dt\,
               \exp [ i \omega' ( t - \mathbf{n} \brho ) ]
        \right|^2
    \right\}.
    }
\end{equation}
This expression is used here in the computations; it is implemented in the open-source code
\textit{jE}~\cite{jE-code}. As for version 1.0.0 used here, for the integrals in
equation~\eref{W} the code uses trapezoidal rule of
integration with a fixed time step. The time step is computed as one-half of the minimal
oscillation period reached on the integration interval $[-t_b, t_b]$ (or two integration intervals
in the case $\varsigma = -1$, $\tau_\perp \gg \tau_\parallel$, see previous subsection). As it is shown
above, numerical error caused by finitness of the integration interval decreases quite slowly,
$\mathcal{I}(t_b, \infty) \propto 1/t_b$. Hence, to obtain proper accuracy one should choose $t_b$
much greater than $t_{rf}$, say $t_b \approx 100 \, t_{rf}$ for accuracy of about $1\%$. At the same
time the oscillation period hence the timestep sharply decreases with time, $T(t_b) \sim 1 / \dot
\phi(t_b) \propto t_b^{-2}$. Thus the resulting number of nodes (time steps) yielding proper
accuracy becomes extremely large. However, the computation of the integrals can be performed on a
much smaller interval, if the artificial attenuation $g$ is added:
\begin{equation}
    \label{math_trick}
    \int_{-\infty}^\infty f(t) \sin[\phi(t)] \, dt \approx \int_{-t_b}^{t_b} g(t) f(t)
    \sin[\phi(t)] \, dt.
\end{equation}
Here $t_b$ should be just several times larger than $t_{rf}$ (in the code $t_b \approx 3
t_{rf}$ is chosen), and the function $g$ should fade smoothly near the boundaries of the integration interval
from $1$ to $0$. Equation~\eref{math_trick} can be easily proven by estimating the difference of
its left-hand-side and its right-hand-side with formula~\eref{convergence_speed}: $(1 - g)fT
\approx 0$ at the point where $g(t)$ just starts to fade as well as at $t_b$. In the code the
following attenuation function is chosen:
\begin{equation}
    g(t) = \frac{1}{4} \left\{1 - \tanh[8 (t / t_b - 0.7)] \right\}
                       \left\{1 + \tanh[8 (t / t_b + 0.7)] \right\},
\end{equation}
which together with the given timestep and the integration interval yields error less than $3\%$ in
comparison with the integrals without $g$ computed on a extremely wide integration interval by a
different numerical method, at least for $\varsigma \tau_\parallel / \tau_\perp \in [-48, 0.8]$.

In the code the integration method described above is used in the function which computes photon
emission probability with equation~\eref{W}. A number of tests is implemented for this function.
For instance, in the classical limit ($\chi \ll 1$) energy radiated per unit frequency interval per
unit solid angle, $d^2I/d\omega d\Omega$, computed numerically, is compared with analytical
results, namely with equation~(14.83) from~\cite{Jackson62}. This test shows accuracy of the code better
than $0.5\%$ for $|\theta| \leq 1 / \gamma$ and $\omega \leq 1.6 \times
\omega_c$. At this point the value of $d^2I/d\omega d\Omega$ is already more than $500$ times lower
than the maximal value of $d^2I/d\omega d\Omega$, thus although the error becomes greater with the
increase of $\omega$ and $\theta$, the whole value of $d^2I/d\omega d\Omega$ can be neglected
there.  Furthermore, a well-known asymptotic behavior of the full radiated energy is also tested
\{namely $cI / 2 \pi R \approx (1 - 55 \sqrt{3} \chi / 16 + 48 \chi^2) P_{cl}$ at $\chi \ll 1$ with
$P_{cl} = 2 e^4 B^2 \gamma^2 / 3 m^2 c^2$ and $cI / 2 \pi R \approx 0.37 \times e^2 m^2 c^4 \chi^{2/3} /
\hbar^2$ at $\chi \gg 1$, see~\cite{Baier98, Berestetskii82}\}.

The radiation spectrum calculated numerically and analytically for $n = 1$ and $\chi = 3$ is
shown in figure~\ref{small_n}(d), where the result of \textit{jE} code is shown with solid red line and
the analytical result with dotted black line. A number of tests also is written in order to demonstrate that the
mass of the emitting particle, the spin term and the refractive index are treated
correctly~\cite{jE-code}.

\section{Possible experimental evidence of vacuum polarization}
\label{estimates}

\subsection{Synchrotron-Cherenkov radiation of electrons}
\label{estimates_electrons}

One can ask for the conditions necessary to modify the well known synchrotron spectrum because of
the vacuum polarization. To answer, one first needs to discuss an expression for the
vacuum index of refraction in a strong field. The refractive index depends on the photon
polarization. For example, in a constant magnetic field $\delta n$ for low-energy photons
is about twice
greater for the polarization perpendicular to the magnetic field, in comparison with the polarization
parallel to the magnetic field. However, the most of the synchrotron photons are polarized
perpendicularly to the magnetic field, and the following expression for the real part of
the vacuum refractive index can be used [see~\cite{Erber66, Ritus70, McDonald86} and references
therein]:
\begin{equation}
	\label{vacuum_refractive_index}
        n(\varkappa) = 1 + \frac{\alpha}{4\pi} \left( \frac{B}{B_S} \right)^2 N(\varkappa),
\end{equation}
with $N(\varkappa)$ is presented in figure~\ref{muons_low_chi}(a), $\varkappa = (\hbar \omega /
mc^2)(B / B_S)$ is the photon analogue of the $\chi$ parameter, and $B$ the (effective) magnetic
field.  The asymptotics of $N(\varkappa)$ are given by:
\begin{equation}
    \label{N}
    N(\varkappa) = 
    \cases{
        \eqalign{
        14/45 \qquad & \mathrm{for} \quad \varkappa \ll 1 \\
        -0.278 \times \varkappa^{-4/3}
              \qquad & \mathrm{for} \quad \varkappa \gg 1
        }
    }
\end{equation}
where $\varkappa =(\hbar \omega / mc^2) (B / B_S)$. As seen from figure~\ref{muons_low_chi},
$\delta n = n - 1$ is
positive for $\varkappa \lesssim 15$ and negative for $\varkappa \gtrsim 15$.

As described in the previous sections, the refractive index influences only the timescale
$\tau_\parallel$ in which the electron becomes out of phase with the wave due to the difference of its
velocity along the wave vector $\mathbf k$ and the wave phase velocity.
Hence the vacuum polarization influences only the linear
term in the phase~\eref{bks_phi}. At the same time the linear term of the phase influences the
integrals in equation~\eref{bks_I} only if the timescale $\tau_\parallel$ is less or about
$\tau_\perp$. In the timescale $\tau_\perp$ the electron becomes out of the phase with the wave due to the trajectory
curvature (because the curvature affects the electron velocity along the wave vector). Summing up, and
taking into account equations~\eref{bks_tau_parallel}, \eref{bks_tau_perp} and \eref{bks_omega_c}, the necessary
condition of the spectrum change at a given photon frequency $\omega$ is the following:
\begin{numcases}{}
    \label{cond1}
        |\delta n(\omega)| \gtrsim 1 / \gamma^2, & \\
    \label{cond2}
        \omega' \gtrsim \omega_c'(\omega). &
\end{numcases}
If condition~\eref{cond1} holds, then $\tau_\parallel$ changes noticeably, and
if~\eref{cond2} is holds too (with $\omega_c'$ computed either with or without vacuum polarization
taken into account), then
the spectrum changes. Here $\omega' = \omega \varepsilon / (\varepsilon - \hbar \omega)$, and $\omega_c'$
is determined by equation~\eref{bks_omega'_c} for a given frequency $\omega$ (note that $\delta n$
depends
on $\omega$). 

It should be noted that the conditions~\eref{cond1} and \eref{cond2} are very weak: if $\delta n$ is, say, $10\%$
of $1/\gamma^2$ it nevertheless can lead to sizable changes in the spectrum. Also, if $\omega'$ is
$10\%$ of $\omega_c'(\omega)$, the spectrum changes noticeably. For instance, for $\chi = 3$ and $\delta n =
0.1 / \gamma^2$, one has $\omega' \approx 0.1 \times \omega_c'(\omega)$ for $\hbar \omega \approx 0.5
\varepsilon$, however, the changes in the spectrum are evident for this frequency, as seen in
figure~\ref{small_n}(d).

For low-energy photons ($\varkappa \ll 1$) equation~\eref{cond1} can be rewritten using the
electron $\chi$ parameter only:
\begin{equation}
    \chi \gtrsim \left( \frac{90 \pi}{7 \alpha} \right)^{1/2} \approx 70.
\end{equation}
However, condition~\eref{cond2} in this case much harder to fulfill: it also can be written in
terms of $\chi$ and yields for $\varkappa = 1$
\begin{equation}
    \label{42}
    \alpha \chi^{2/3} \gtrsim \frac{3^{2/3} 45 \pi}{7} \approx 42,
\end{equation}
hence $\chi \gtrsim 4 \times 10^5$. One can note that the result~\eref{42} is far beyond the
conjectured threshold of the perturbative QED breakdown~\cite{Fedotov17}, $\alpha \chi^{2/3} \gtrsim
1$.  Therefore, the BKS approach used here and the expression for the refractive index~\eref{N} are
hardly valid if $\alpha \chi^{2/3} \gtrsim 1$ and even more so for $\alpha \chi^{2/3} \gtrsim 42$.
Thus the considered theory predicts no change in the synchrotron spectrum in the region of the
perturbative QED.

For high-energy photons ($\varkappa \gg 1$) equation~\eref{cond1} yields $\chi \gtrsim 80 \,
\varkappa^{2/3}$, and equation~\eref{cond2} yields $\alpha \chi^{2/3} \gtrsim 20 \, \varkappa (\chi
- \varkappa) / \chi$. To fit the latter to the region of perturbative QED applicability, one can
try $\chi - \varkappa \ll \chi$, however, the former condition in this case yields $\chi^{1/3}
\gtrsim 80$ hence $\chi \gtrsim 5 \times 10^5$ which is again beyond the region of perturbative
QED. Therefore, the evidence of vacuum polarization in synchrotron spectrum for high-energy photons
is also unreachable. The estimates above are in agreement with the results of
reference~\cite{Erber76} [see equations~(8.8e) and (8.11) therein], which, however, do not take
radiation recoil into account and do not discuss the region of the perturbative QED applicability.

\subsection{Synchrotron-Cherenkov radiation of muons}
\label{estimates_muons}

The effect of the vacuum polarization on the synchrotron spectrum can be enhanced if heavy charged
particles are used instead of electrons. For definiteness, and because of the recent progress in
their acceleration technique~\cite{MICE20}, muons are considered here. The advantage of using muons is a
two-fold. First, their big mass, $m_\mu \approx 207 \, m$, yields much greater curvature
radius hence much greater timescale $\tau_\perp$ than that for the electrons. For a given photon frequency
this makes synchrotron spectrum much more sensible to the longitudinal synchronism between the
particle and the emitted wave, i.e. to $\tau_\parallel$ which, opposite to $\tau_\perp$, depends on
the refractive index. Second, high mass makes the critical frequency significantly lower, hence a
more sizable part of the spectrum lies in the low-frequency region $\varkappa \lesssim 1$, in which
$\delta n$ for the vacuum refractive index is maximal.

The classical critical frequency for muons is
\begin{equation}
    \label{omega_c_mu}
    \omega_{c, \mu} = 3 \omega_B \gamma^2 m / m_\mu,
\end{equation}
that gives the ratio of the photon energy to the muon energy: $\hbar \omega_{c, \mu} /
\varepsilon_\mu = 3 \chi (m / m_\mu)^2$ (with $\chi = \gamma B / B_S$ the same as for
electrons).  Therefore, this ratio is small, $\hbar \omega_{c, \mu} / \varepsilon_\mu \ll 1$, up to
$\chi \sim 10^4$, and it is reasonable to neglect the radiation recoil for muons. In this case
the conditions sufficient for the synchrotron spectrum to be noticeably modified due to vacuum polarization are
the following:
\begin{numcases}{}
    \label{mu_cond1}
        |\delta n(\omega)| \gtrsim 1 / \gamma^2, & \\
    \label{mu_cond2}
        \omega \gtrsim \frac{\omega_{c,\mu}}{\left| 1 - 2 \delta n(\omega) \gamma^2 \right|^{3/2}}.
        &
\end{numcases}
Obviously, these conditions can be derived similarly to equations~\eref{cond1} and \eref{cond2}.

\begin{figure*}
	\includegraphics{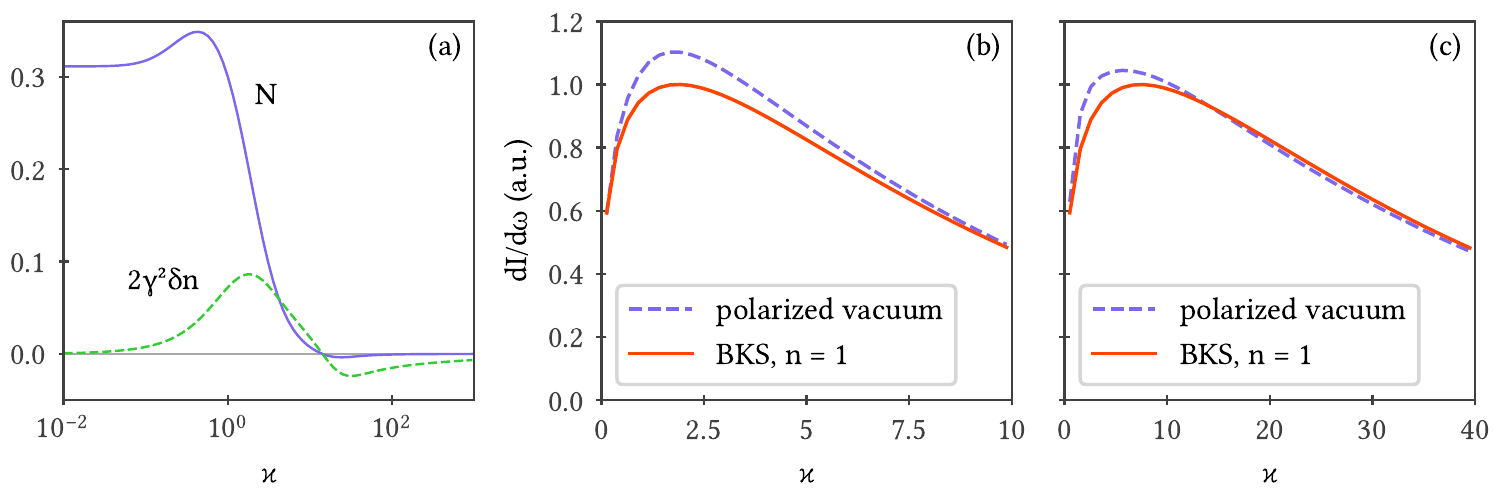}
    \caption{\label{muons_low_chi} (a, solid blue line) Function $N(\varkappa)$ which determines
    the real part of the vacuum refractive index~\eref{vacuum_refractive_index}, with $\varkappa =
    (\hbar \omega / mc^2) (B / B_S)$, and (a, dashed green line) combination $2 \gamma^2 \delta
    n(\omega_c/3)$ with $\omega_c$ computed as classical critical frequency for muons from the parameter
    $\varkappa$. Note that the relative change of $\tau_\parallel$ caused by vacuum polarization
    for $\theta = 0$ is approximately equal to $2 \gamma^2 \delta n$, if this combination is small.
    Muon radiation spectra for (b) $\chi = 30$ and (c) $\chi = 60$ (where $\chi = \gamma B /
    B_S$), with (dashed blue line) vacuum refractive index taken into account and (red line) for
    $n = 1$.}
\end{figure*}

As a starting point, one can consider $\omega \sim \omega_c$ that ensures fulfillment of
condition~\eref{mu_cond2}. By virtue of equation~\eref{bks_tau_parallel},
the difference between $\tau_\parallel$ with $\delta n \neq 0$ taken into account, and
$\tau_\parallel$ with $\delta n = 0$ ($\tau_{\parallel, 0}$), is determined by $2 \gamma^2 \delta
n$, if this quantity is small:
\begin{equation}
    \frac{\tau_\parallel - \tau_{\parallel, 0}}{\tau_\parallel} \approx 2 \gamma^2 \delta n,
\end{equation}
where $\theta = 0$ is assumed. One can note that this quantity depends on $\chi$ and
$\varkappa$ only [see equation~\eref{vacuum_refractive_index}]. For $\omega = \omega_c / 3$ (which
corresponds to the maximum of the spectrum better than $\omega_c$ itself) the parameters $\chi$ and
$\varkappa$ become related, $\chi^2 = \varkappa m_\mu / m$.  Thus $2 \gamma^2 \delta n$ can be
expressed in terms of $\varkappa$ only, and $2 \gamma^2 \delta n$ as function of $\varkappa$ is
plotted as dashed green line in figure~\ref{muons_low_chi}(a).  It is seen from
figure~\ref{muons_low_chi}(a) that the vacuum polarization causes maximal change in $\tau_\parallel$ of
about $10\%$ for $\varkappa \approx 2$ ($\chi \approx 20$). Further increase of the parameter
$\chi$ (hence $\varkappa$) leads to the decrease of the change of $\tau_\parallel$ at this photon
frequency.

The estimates above demonstrate that relatively small value of $\chi$ is enough to see the change
in the synchrotron spectrum caused by the vacuum polarization, that agrees well with the numerical
results.  Figures~\ref{muons_low_chi}(b, c) demonstrate the radiation spectrum for $\chi = 30$ and
$\chi = 60$, respectively, computed with \textit{jE} code (value $B / B_S = 0.01$ is used,
hovewer, the shapes of the resulting spectra almost do not depend on it). At $\chi = 30$ the spectrum
maximum becomes $10\%$ higher thanks to the vacuum polarization, and at higher $\chi$ changes in
the spectrum occurs at frequencies lower than the frequency of the spectrum maximum. At the same time,
for photon energies corresponding to $\varkappa \gtrsim 15$, the change in the refractive index
becomes negative that causes slight quenching in the synchrotron spectrum.

Returning to the conditions~\eref{mu_cond1} and \eref{mu_cond2}, one can consider $\chi \gtrsim 70$
that ensures the fulfillment of the first of them. Then, similarly to the previous section, the
low-energy part of the spectrum ($\varkappa \lesssim 1$) can be considered. In this case
condition~\eref{mu_cond2} yields
\begin{equation}
    \label{mu_42}
    \alpha \chi^{2/3} \gtrsim 42 \times \left( \frac{m}{m_\mu} \right)^{2/3} \approx 1
\end{equation}
[compare this with equation~\eref{42}]. Therefore, the change in the low-energy part of the
synchrotron spectrum can be pronounced only near the conjectured breakdown threshold of the perturbative
QED.

\begin{figure*}
	\includegraphics{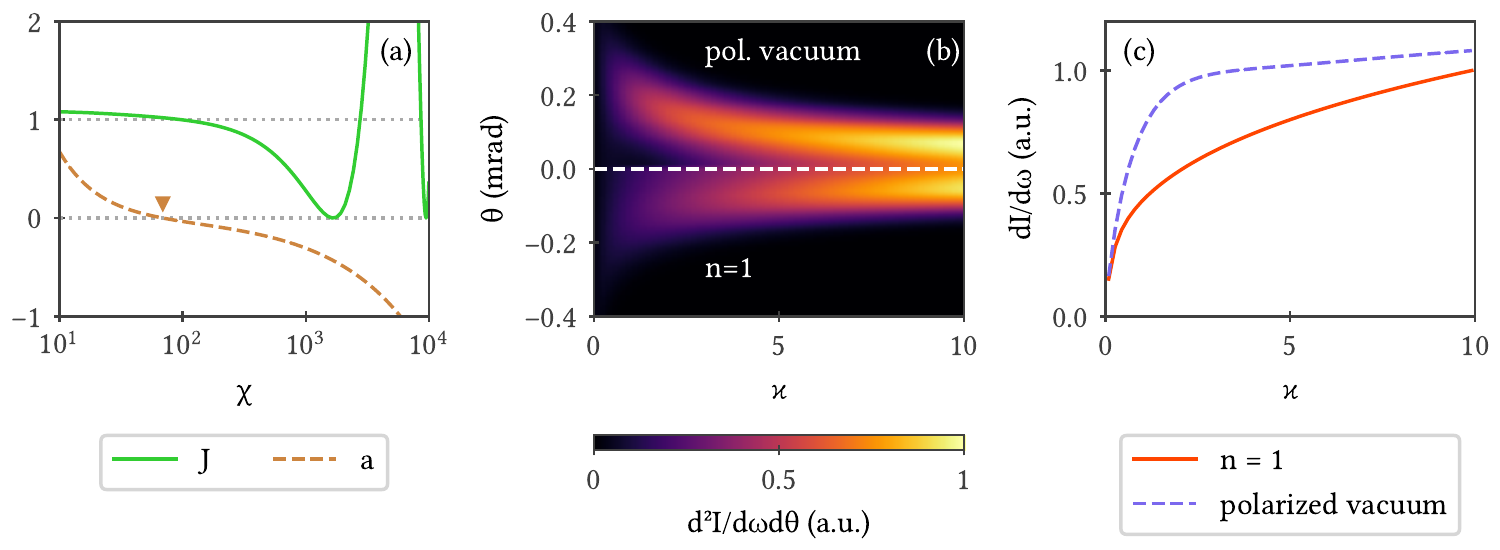}
    \caption{\label{muons} (a, green line) The ratio $J$ of the energy radiated by a muon per unit
    frequency per unit $\theta$ computed for vacuum refractive index, to the same quantity computed
    for $n = 1$. (a, brown line) The ratio $a = \varsigma \tau_\perp / \tau_\parallel$; the arrow
    marks the point at which $a = 0$ ($\chi \approx 70$).  Values for the both lines are computed for
    $\theta = 0$ and value of $\omega$ providing $\varkappa = 2$.  (b) $d^2 I / d\omega d\theta$ and (c)
    muon radiation spectrum for $\chi = 800$ with (b: upper half, c: dashed blue line) vacuum
    polarisation taken into account and (b: lower half, c: solid red line) for $n \equiv 1$.}
\end{figure*}

Figure~\ref{muons} demonstrates the low-energy part of the radiation spectrum of muons for $B / B_S
= 0.01$ (note that the subplots almost do not depend on this value). In figure~\ref{muons}(a) the
ratio
\begin{equation}
    J = \left. \left[ \frac{d^2 I}{d\omega d\theta} \right] \middle/ \left[ \frac{d^2 I}{d\omega
    d\theta} \right]_{\delta n = 0} \right. ,
\end{equation}
computed for $\theta = 0$ and value of $\omega$ providing $\varkappa = 2$, is shown with green solid line.
According with the estimate~\eref{mu_42}, value of $J$ differs noticeable from unity for $\chi \sim
10^3$. The dashed brown line shows the ratio $a = \varsigma \tau_\perp / \tau_\parallel$ for the
given frequency which corresponds to $\varkappa = 2$. It is seen from the expression for the
classical critical frequency [equation~\eref{omega_c_mu}] that
\begin{equation}
    \frac{ \hbar \omega_{c, \mu} }{ mc^2 } \frac{B}{B_S} = \frac{3 \chi^2 m}{m_\mu}.
\end{equation}
Hence, for $\chi \gtrsim 10$ the frequency which provides $\varkappa = 2$ is lower than the critical frequency.
Thus, one expects $\tau_\perp \lesssim \tau_\parallel$ here, for $n = 1$.
However, at $\chi \sim 10^3$ the timescales $\tau_\perp$ and
$\tau_\parallel$ becomes of the same order thanks to the vacuum polarization. At even higher values of $\chi$, $\tau_\parallel$
becomes much smaller than $\tau_\perp$ that together with $\varsigma = -1$ leads to the interference
patterns similar to that shown in figure~\ref{sinus2}(b). In figure~\ref{muons}(a) this is seen as
the oscillations of $J$ at high values of $\chi$.

Figure~\ref{muons}(b) shows the radiated energy per unit frequency and per unit angle $\theta$ for
the vacuum refractive index taken into account (upper half) and $\delta n = 0$ (lower half), for $\chi
= 800$ ($B / B_S = 0.01$ and $\gamma = 8 \times 10^4$). Figure~\ref{muons}(c) shows the energy
spectra which correspond to the distributions in figure~\ref{muons}(b). Although the difference
between dashed and solid curves in figure~\ref{muons}(c) is dramatic, it is hardly fit as possible
experimental evidence of the vacuum polarization. First, it rises only at high $\chi$ values, where
some other high-order terms of QED can give even bigger contribution. Second, the change in
the spectrum occurs only for frequences for which $\varkappa \lesssim 10$, which is much lower than for
the critical frequency, hence this difference occurs for a small fraction of the photons.
Therefore, photon emission by muons with $\chi \approx 30$ is still the most promising probe
for the vacuum polarization effect in the radiation spectrum.

One more interesting prospect of QED study should be noted regarding the photon emission by muons.
Let muons and electrons are of the same Lorentz factor $\gamma$. For the electrons the energy of
the emitted photons in the regime of $\chi \gg 1$ is limited due to the recoil effect by $mc^2
\gamma$. Despite higher curvature radius, for the muons the photon energy can be much higher,
because the recoil effect for them is negligible. Definitely, one can reach $\alpha
\varkappa^{2/3} \sim 1$ for photons emitted by the muons already at $\chi \sim 300$, for which
$\alpha \chi^{2/3} \sim 0.3$. This potentially opens perspectives to reach the non-perturbative
QED~\cite{Fedotov17, Yakimenko19, Baumann19, Blackburn19b, Piazza19a} in future experiments.

\section{Conclusion}
\label{conclusion}

The general formula which describes the photon emission by an ultrarelativistic electron in a
strong magnetic field can be found in the framework of the quasiclassical theory of Baier and
Katkov~\cite{Baier98}. Baier--Katkov formula can be extended to the case of a constant non-unity
refractive index $n$, $|n - 1| \ll 1$ [see equation~\eref{bks_I}]. From this, one can find photon
emission probability that generalizes both the synchrotron and the Cherenkov emission, and takes into
account photon recoil and spin flips. The obtained expression clearly shows that the emission probability
is not the sum of the synchrotron emission probability and the Cherenkov emission probability.
Hence, the photon emission occurs in the synergic (cooperative) synchrotron-Cherenkov radiation
process.

The electron motion along its curved trajectory prevents the pure Cherenkov radiation.  The
trajectory curvature determines the radiation formation time for the synchrotron-Cherenkov
radiation [see equation~\eref{t_rf}], which is much shorter than that for the pure Cherenkov
radiation (which can be extremely large in the case of the Cherenkov synchronism, ${\mathbf n
\bbeta} = 1$).  Furthermore, the radiation formation time for the synchrotron-Cherenkov radiation
is less or about of that for the pure synchrotron emission.

The photon emission probability is determined not only by the radiation formation time, and the
probability can be either greater or less for the synchrotron-Cherenkov radiation ($n \neq 1$) than
that for the synchrotron one ($n = 1$). The radiation spectrum is sensible to $\delta n = n - 1$ in
the both cases, $\delta n > 0$ and $\delta n < 0$, however, the changes in the spectrum occur first
for frequencies which are higher than the critical frequency $\omega_c$ [see
equation~\eref{bks_omega_c}]. If the Cherenkov condition holds, $v > c / n$, the overall emitted
energy can be much higher than in the case $n = 1$.  For numerical simulations of the
synchrotron-Cherenkov spectrum the open source code \textit{jE} is implemented~\cite{jE-code}, and the
numerical results are in a good agreement with the analytical predictions.

One can use formulas for the refractive index of vacuum polarized by a strong external magnetic
field \{see reference~\cite{Tsai75} and references therein\}, in order to find how the synchrotron
spectrum is modified due to vacuum polarization. The estimates and numerical simulations
demonstrate that in the framework of the considered model the changes in the spectrum emitted by
electrons becomes noticeable far beyond the Cherenkov threshold $v = c / n$ and even far beyond the
conjectured breakdown of the perturbative QED $\alpha \chi^{2/3} \sim 1$. The cause of this is
that the vacuum refractive index depend on the photon frequency, and for the photon energies
greater or about $\hbar \omega_c$ (for which the spectrum modification is expected first)
$\delta n$ is negative and $|\delta n|$ is very small. Moreover, the critical frequency for
electrons with $\chi \gg 1$ is very close to the electron energy.

Muons have much larger curvature radius of the trajectory in a strong field than the electrons, if
the muons and the electrons are of the same Lorentz factor. This makes the radiation spectrum of
the muons much more sensible to the refractive index than that of the electrons. Opposite to the
electrons with $\chi \gg 1$ for which the critical frequency always yields $(\hbar \omega_c /
mc^2)(B / B_S) \gg 1$, for the muons this is not the case. E.g., for $\chi = 30$ the maximum of the
spectrum corresponds to $\varkappa = (\hbar \omega / mc^2)(B / B_S) \approx 2$ which is favourable
for the vacuum refractive index. The radiation spectrum is enhanced up to $10\%$ in this case
thanks to the vacuum polarization [see figure~\ref{muons_low_chi}(b)]. From the point of view of
possible experiments, the muons with $\chi \approx 30$ probable are the most promising tool to
probe the influence of the vacuum polarization on the synchrotron spectrum.

Regarding possible experiments using laser pulses and muon accelerators, a simple head-on collision
geometry with single laser pulse can be considered. The expression for the vacuum refractive index
for the emitted photons in this case is quite close to that for a constant magnetic
field~\cite{Ritus70, McDonald86}. However, the results of this paper can not be applied directly to
the laser field. First, as the most of the emitted photons have $\varkappa \gtrsim 1$, the
pair photoproduction and the photon emission by the secondary electrons and positrons should be
taken into account. Second, $\chi = 30$ will be reached rather at $a_0 = eE_0 / mc \omega_L \sim
m_\mu / m = 207$ (with $E_0$ and $\omega_L$ the electric field amplitude and the photon frequency
of the laser pulse, respectively), e.g. for $a_0 = 800$
and $\gamma = 2 \times 10^4$ (for $\hbar \omega_L = 1 \; \mathrm{eV}$). For such values of $a_0$
the local constant field approximation applied here should be used with caution~\cite{Piazza19}.
This constraint becomes even more pronounced for protons, for which the dipole-Cherenkov radiation
should be considered rather than the synchrotron-Cherenkov radiation. Third, in the linearly
polarized laser the refractive index is not uniform (however, it probably can be considered
uniform on a scale of the radiation formation length). Therefore, the realistic proposal for
probing vacuum polarization with synchrotron emission of heavy charged particles in laser fields needs further
investigations.

\ack

We thank A.~M.~Fedotov and A.~A.~Mironov for fruitful conversations and comprehensive references.

This research is supported by the Russian Science Foundation through Grant No. 18-72-00121.

\bibliographystyle{iopart-num}
\bibliography{main}

\end{document}